\documentclass[floats,floatfix,showpacs,preprintnumbers,amssymb,prd,twocolumn,superscriptaddress,nofootinbib,nolongbibliography,reprint]{revtex4-1}

\usepackage{amssymb,amsmath,verbatim,mathtools,needspace,enumitem,etoolbox,graphicx,physics,microtype,afterpage,xspace,tabularx,lmodern,multirow}
\usepackage[utf8]{inputenc} 
\usepackage{graphicx,xcolor,overpic,mathtools}
\usepackage[colorlinks]{hyperref}
\hypersetup{ 
	colorlinks=true, 
	linkcolor=black, 
	filecolor=black, 
	citecolor = blue,       
	urlcolor=blue, 
} 
\newcommand{\beq}{\begin{equation}}
\newcommand{\eeq}{\end{equation}}

\usepackage{upgreek}
\usepackage{csquotes}
\usepackage{verbatim}
\usepackage{tensor}

\def\beq{\begin{equation}} 
\def\eeq{\end{equation}}
\def\beqa{\begin{eqnarray}}
\def\eeqa{\end{eqnarray}}

\PassOptionsToPackage{normalem}{ulem}
\usepackage{ulem} 
\usepackage{sidenotes}
\usepackage{bm}
\usepackage{url}
\usepackage{xfrac}
\usepackage[makeroom]{cancel}
\usepackage{tensor}
\usepackage{mathrsfs}
\usepackage{slashed}
\usepackage{tikz}
\usepackage[mode=buildnew]{standalone}
\usepackage{bbm}
\usepackage{mathtools} 
\usepackage{cleveref}
\crefname{section}{Sec.}{sec.}
\crefname{figure}{Fig.}{Fig.}
\Crefname{section}{Section}{Sections}
\usepackage{caption,subcaption}
\DeclareMathOperator{\MeV}{MeV}

\DeclareMathOperator{\fm}{fm}

\NewDocumentCommand{\evat}{sO{\bigg}mm}{%
  \IfBooleanTF{#1}
   {\mleft. #3 \mright|_{#4}}
   {#3#2|_{#4}}%
}

\newcommand{\prlsec}[1]{\noindent {\bf \em #1. }}

\begin{document}

\count\footins = 1000 
\title[
]{Electroweak form factors of large nuclei as BPS skyrmions}
\author{Alberte Xosé López Freire}
\email{albertexose.lopez.freire@usc.es}
\author{Christoph Adam}
\email{christoph.adam@usc.es}
\affiliation{%
Departamento de F\'isica de Part\'iculas, Universidad de Santiago de Compostela E-15782 Santiago de Compostela, Spain
}
\affiliation{Instituto
Galego de F\'isica de Altas Enerxias (IGFAE) E-15782 Santiago de Compostela, Spain
}
\author{Alberto García Martín-Caro}
\email{alberto.garcia.martin-caro@uvigo.es}
\affiliation{EHU Quantum Center and Department of Physics, University of the Basque Country UPV/EHU, Bilbao, Spain}
\affiliation{Instituto de Física e Ciencias Aeroespaciais (IFCAE), Universidade de Vigo. 32004 Ourense, Spain}
\author{Diego González Díaz}
\email{diego.gonzalez.diaz@usc.es}
\affiliation{%
Departamento de F\'isica de Part\'iculas, Universidad de Santiago de Compostela E-15782 Santiago de Compostela, Spain
}

\date[ Date: ]{\today}
\begin{abstract} 
We employ the Bogomolnyi-Prasad-Sommerfield (BPS) Skyrme model within the framework of semi-classical quantization as an effective model to compute both the electromagnetic and neutral current form factors for heavy nuclei. Our results show excellent agreement with the experimental data for low- to moderate momentum transfer. Further, we present an analytic expression of the neutral current form factor for generic nuclei, expressed as a power series in the momentum transfer.
Our method provides an alternative to existing phenomenological approaches  which, after fitting just one global radial parameter, allows for a surprisingly precise determination of the electroweak form factors at low momentum transfer for all heavy nuclei. Such a simple and robust description is particularly relevant for precision neutrino experiments, because it allows for a certain control over model-dependent systematics, which is essential for probing physics beyond the Standard Model.
\end{abstract}
\maketitle

\prlsec{Introduction}
While the electromagnetic charge distribution of atomic nuclei has been extensively studied through elastic electron scattering experiments, determining the neutron density distribution remains a significant challenge. Enhanced measurements of neutron densities would have profound implications for the equation of state of neutron-rich matter, which is crucial for understanding the structure and evolution of neutron stars \cite{Thiel:2019tkm, Reed:2021nqk}, as well as for deriving constraints of new physics beyond the Standard Model \cite{Barranco:2005yy, Dutta:2019eml}. Although precise experimental data exist for observables sensitive to neutron distributions--such as nuclear dipole polarizability \cite{Tamii:2011pv,Hashimoto:2015ema}--hadronic probes introduce model-dependent uncertainties that require careful analysis (see, e.g., Ref. \cite{Thiel:2019tkm}).

In contrast, electroweak processes such as parity-violating electron scattering (PVES) \cite{Prescott:1978tm,Souder:2015mlu} and coherent elastic neutrino-nucleus scattering (CE$\nu$NS) \cite{Freedman:1973yd} have long been regarded as cleaner and more direct methods for probing neutron densities \cite{Ciuffoli:2018qem,Ruso:2022qes}. Despite their experimental complexity, both techniques have seen significant advances in recent years \cite{COHERENT:2017ipa, COHERENT:2019iyj,COHERENT:2024axu,Abrahamyan:2012gp,CREX:2022kgg,CONUS}. These efforts provide valuable insights into nuclear structure by enabling the determination of the weak form factor\footnote{In the case of neutrino probes, this is true at
least at low momentum transfers where the process remains coherent.}, which directly constrains the neutron density distribution within nuclei \cite{Cadeddu:2019eta,Papoulias:2019lfi}.

Since no precision measurements of neutron density distributions of nuclei are available at the moment, the weak nuclear form factor (FF) has to be modelled in order to evaluate experimental cross
sections and event rates. Traditionally, the weak FF has been modelled using phenomenological approaches based on empirical fits to elastic electron scattering data, such as the Klein–Nystrand (KN) FF \cite{Klein:1999qj} or the Helm FF \cite{Helm:1956zz}, both widely used in the CE$\nu$NS community \cite{Baxter:2019mcx}. 
However, in recent years a theoretical effort has been made to develop actual nuclear structure calculations of ground states and density distributions through microscopic nuclear physics approaches such as density
functional theory \cite{Patton:2012jr}, relativistic mean field methods \cite{Yang:2019pbx}, coupled–cluster theory from first principles \cite{Payne:2019wvy}, shell-model
calculations \cite{AbdelKhaleq:2024hir}, Hartree–Fock plus Bardeen–Cooper–Schrieffer model \cite{Co:2020gwl} as well as effective
field theory approaches \cite{Tomalak:2020zfh,PhysRevD.102.074018},
that provide a more accurate description of the weak FF. However, these highly sophisticated methods often require long computations to determine the properties of individual nuclei or the fitting of a large number of free parameters that reduce their predictive power.

In this work, we present the derivation of electroweak FF from a very simple effective model of low energy nuclear physics, a variation of the well known Skyrme model \cite{skyrme,manton1}.


The Skyrme model is a nonlinear field theory of pions which describes nucleons, nuclei, and nuclear matter in terms of appropriately quantized topological soliton solutions (``skyrmions'') \cite{adkins1983static,adkins1984skyrme,Carson:1991fu,battye2009light}. Chiral symmetry is realized non-linearly in this model, by pion fields taking values in the group manifold SU(2). 
A general Skyrme field can be expressed as
\begin{equation} \label{ec:defU}
    U(x) = \sigma(x) \mathbbm{1} + i \boldsymbol{\pi}(x) \cdot \boldsymbol{\tau},
\end{equation}
where $x=(\mathbf{x},t)$, $\boldsymbol{\tau} $ is the vector of Pauli matrices, $\boldsymbol{\pi}=(\pi_1,\pi_2,\pi_3) , \pi_i \in \mathbb{R}$ are the pion fields and $\sigma(x) \in \mathbb{R}$ is an auxiliary field obeying
 $   \sigma^2 + \boldsymbol{\pi} \cdot \boldsymbol{\pi} = 1.$
The nontrivial topology of the field space endows each finite energy field configuration $U$ - and, in particular, each skyrmion - with an invariant, integer topological degree,
\begin{equation} \label{ec:topchargue}
    B = \int \mathcal{B}^0 d^3 x\,  ;\; \;  \mathcal{B}^\mu =  \frac{1}{24 \pi^2} \varepsilon^{\mu\nu \rho \sigma} \Tr{L_\nu L_\rho  L_\sigma}
\end{equation}
($L_\mu = U^{-1}\partial_\mu U$), which is identified with the baryon number.

The model originally introduced by Skyrme is defined by the Lagrangian density\footnote{The index $i$ in $\mathcal{L}_i$ denotes the power of first derivatives of this term.}
\begin{equation}
   \mathcal{L}_{S} = \mathcal{L}_2 + \mathcal{L}_4\equiv -\frac{f_\pi^2 }{16} \Tr (L_\mu L^\mu)+\frac{1}{32e^2} \Tr(\left[L_\mu,L_\nu \right]^2)
\end{equation}
where the quadratic term
is dictated by current algebra arguments, and 
Skyrme introduced the quartic term
in order to bypass the scaling instability of $\mathcal{L}_2$ \cite{Derrick:1964ww} such that solitonic solutions can exist. 

Our starting point is the generalized Skyrme model (first considered in \cite{Jackson:1985yz}),
\begin{equation}
    \mathcal{L}_{GSM} =  \mathcal{L}_S + \mathcal{L}_{BPS}, \;\;  \mathcal{L}_{BPS} \equiv \mathcal{L}_0 +\mathcal{L}_6
\end{equation}
where 
\begin{equation}
    \mathcal{L}_0 \equiv -\mu^2 V (\Tr U)
\end{equation}
is a potential term which explicitly breaks chiral symmetry down to isospin symmetry and serves, e.g., to give a mass to the pion field. Finally, 
\begin{equation} \label{ec:L6}
    \mathcal{L}_6 \equiv -\lambda^2 \pi^4 \mathcal{B}_\mu \mathcal{B}^\mu , 
\end{equation}
is proportional to the square of the baryon current. This is the only further Poincare-invariant term not more than quadratic in time derivatives that can be added to the model. 

Conceptually, the model $\mathcal{L}_{GSM}$ constitutes a candidate theory for nuclear and hadron physics in terms of pions, where baryons and nuclei are realized as topological solitons. However, as an all-encompassing model of nuclei, right now the model lacks precision and is not competitive with the standard methods of nuclear physics.\footnote{But there are strong indications that the inclusion of further vector mesons into the model significantly improves its ability to precisely describe nuclear matter \cite{Naya:2018kyi,Harland:2024dca,Huidobro:2024hgg}.}
On the other hand, the Skyrme model performs much better once more specific problems are considered. It allows, e.g., to predict the rotational excitations of the ground and Hoyle states of $^{12}$C \cite{Lau:2014baa} or a plethora of several dozens of excitations of $^{16}$O \cite{Halcrow:2016spb} with surprising precision. It also provides a good description for high density nuclear matter \cite{Adam:2023cee} and, as a consequence, for the inner core of neutron stars \cite{Adam:2020yfv}. In the high-density regime, the sixth order term \eqref{ec:L6}, which can be related to the $\omega$-meson repulsion, is vital for a correct description of nuclear matter \cite{Adam:2015lra}.

This last observation motivates the introduction \cite{Adam:2010fg} and the study \cite{Adam:2010ds} of the BPS submodel $\mathcal{L}_{BPS} = \mathcal{L}_0 +\mathcal{L}_6$. The addition of $\mathcal{L}_0$ is required to avoid the scaling instability. This submodel, which describes a perfect fluid, is a drastic simplification that eliminates, e.g., propagating pion d.o.f. Nevertheless, it reproduces certain bulk properties of large nuclei and nuclear matter surprisingly well. Like $\mathcal{L}_S$ \cite{Faddeev:1976pg} and $\mathcal{L}_{GSM}$ \cite{Adam:2013tga}, the static energy of $\mathcal{L}_{BPS}$ has a topological lower energy bound \cite{Adam:2010fg}.
Unlike the first two models, the BPS submodel supports infinitely many static BPS solutions that saturate the bound. As a consequence, classical soliton solutions imply zero binding energies for the nuclei they describe, and small, realistic binding energies of large nuclei are achieved by small quantum and Coulomb corrections \cite{Adam:2013wya}. 

Skyrme models also provide a natural arena for the calculation of local charge densities and their associated nuclear form factors, both in the electroweak \cite{Braaten:1988bn,Braaten:1986md,Karliner:2015qoa,GarciaMartin-Caro:2023nty}, and, more recently, the gravitational  \cite{Cebulla:2007ei,GarciaMartin-Caro:2023klo,GarciaMartin-Caro:2023toa} case. 
The semi-classical nature of nuclei in such models allows for a simple calculation of current density operators and their expectation values for general nuclei, and the form factors are then found by a simple Fourier transform. We will find that both electromagnetic and weak neutral form factors for large nuclei can be calculated semi-analytically in the BPS Skyrme model, leading to stunning agreement with experimental data.
\\

\prlsec{The BPS Skyrme Model}
The static energy of the BPS submodel (here, $(1/2)\Tr U = \sigma \equiv \cos \xi $),
\begin{eqnarray}
    E_{BPS} &=& \int \left(\lambda^2 \pi^2 \mathcal{B}_0^2 + \mu^2 V (\xi)\right) d^3x  \\
    &=& \int \left(\lambda \pi \mathcal{B}_0 - \mu \sqrt{V}\right)^2 d^3x \,+\, 
     2 \pi^2 \mu \lambda |B| \langle \sqrt{V}\rangle  \nonumber
\end{eqnarray}
(where $\langle \sqrt{V}\rangle $ is the field-space average of $\sqrt{V}$), can be expressed as the sum of a non-negative term plus a topological lower bound, as is common for BPS theories. 
Solutions that saturate the bound must obey the simpler first-order BPS equation $\lambda \pi \mathcal{B}_0 = \mu \sqrt{V}$. Further, this equation is compatible with an axially symmetric ansatz leading to spherically symmetric energy and baryon densities. Indeed, introducing $\sigma = \cos \xi$, $\boldsymbol{\pi} = \sin \xi \, \boldsymbol{n}$, where $\boldsymbol{n}$ is a three-component unit vector, and inserting the ansatz $\xi = \xi (r)$ and
\begin{equation} \label{ec:nBPS2}
    \boldsymbol{n} (\theta, \phi) = \left(\cos (B \phi ) \sin \theta, \sin (B \phi ) \sin \theta , \cos \theta \right)
\end{equation}
in spherical polar coordinates, results in a first-order ODE for $\xi (r)$. 

Up to this point, all arguments are valid for arbitrary non-negative potentials $V(\xi)$ which take their vacuum value at $\xi =0$, i.e.,  $V(\xi =0)= 0$, such that chiral symmetry is explicitly broken down to isospin symmetry. The correct physical choice of this potential should come from a comparison to the underlying theory of QCD, as is done, e.g., in chiral perturbation theory (ChPT). In ChPT, all terms allowed by chiral symmetry and the particular pattern of its breaking by nonzero light quark masses are introduced and then ordered in a power series in derivatives and quark masses. In the Skyrme model (or, in fact, in any topological soliton model), the spatial derivatives cannot be assumed to be small, because this would directly contradict the Derrick scaling condition for the existence of solitons. On the other hand, the light quark masses are still small in comparison to the QCD scale, and the leading potential term should correspond to the linear quark mass term and reproduce the Gell-Mann--Oakes--Renner relation $f_\pi^2 m_\pi^2 \sim (m_u + m_d)\langle \bar \psi \psi \rangle$. This condition uniquely selects the so-called pion mass potential $ V(\xi) = 1 - \, \cos \xi $ as the leading term. Further terms can be added to the potential, but should be parametrically small and only give small contributions. For simplicity, we shall restrict to the leading term, i.e., the pion mass potential. 

For this potential, the BPS equation can be integrated and leads to an exact solution:
\begin{equation}
        \xi(r) =    \displaystyle 2 \arccos (\frac{r}{R_B})\,\, {\rm for}\,\,  r \in\left[0, R_B\right],\,\, \xi(r)=0 \,\, {\rm otherwise},
        \label{eq:exactsol}
\end{equation}
with $R_B=\sqrt{2}\sqrt[3]{\frac{B \lambda }{\mu }}$. Hence, the solution is of compacton type, with $R_B$ the compacton radius. Further, the BPS energy is linear in $B$, and the (compacton, RMS, etc.) radii behave like $R\sim B^{1/3}$, which is a reasonable order zero approximation to physical nuclei. As said, more realistic binding energies can be achieved by small (quantum and Coulomb) corrections. 

We remark that our results do not strongly depend on this compacton nature. E.g., if we glue an exponential tail to solution \eqref{eq:exactsol} sufficiently close to $r=R_B$  - corresponding to the addition to a further, subleading term to the potential - then the resulting form factors remain essentially unchanged for not too large momentum transfer.
\\

\prlsec{EW conserved currents and form factors}
A simple way to identify the  electroweak (EW) currents in the Skyrme model is {\em i)}, to introduce a covariant derivative $D_\mu$ into its Lagrangian density which ensures local $SU(2)_L \times U(1)_Y$ invariance \cite{Braaten:1986md},
\begin{equation}
    D_\mu U = \partial_\mu U + \frac{ie}{2\sin \theta_W} W^a_\mu \tau_a U - \frac{ie}{2 \cos \theta_W} Y_\mu U \tau_3,
\end{equation}
and, {\em ii)}, to decompose the gauge fields $W^a_\mu$ and $Y_\mu$ into the physical vector bosons
\begin{eqnarray} 
  \hspace*{-0.3cm}  W_\mu^1 = \frac{1}{\sqrt{2}} (W_\mu^+ + W_\mu^-), && W_\mu^2 = \frac{i}{\sqrt{2}} (W_\mu^+ - W_\mu^-), \nonumber \\
\label{ec:W3G}
    W^3_\mu = c_W Z_\mu + s_W A_\mu, && Y_\mu = -s_W Z_\mu + c_W A_\mu, 
\end{eqnarray}
where $c_W \equiv \cos \theta_W, ~s_W \equiv \sin \theta_W$ and $\theta_W$ is the Weinberg angle, with  $s_W^2=0.223$ \cite{CODATA}.
The electroweak currents are now given by the terms that couple to each bosonic field. At this stage, it is convenient to recast these expressions in a more compact form, in terms of the vector and axial-vector currents. 
The derivative terms in the general Skyrme model are invariant separately under left ($U \to VU$) and right ($U \to UW$) chiral transformations. It is useful to rewrite them in terms of
the vector and axial-vector transformations 
\begin{equation}
    \begin{array}{lll}
        V: \quad U & \rightarrow & W^\dagger UW  , ~W \in SU(2) \\
           A: \quad U & \rightarrow & WUW   .
    \end{array}
\end{equation}
In the BPS Skyrme model, their associated Noether current densities are
\begin{equation} \label{ec:Jclas}
      \displaystyle J^{\alpha}_{k, V(A)}
       =  \mp \displaystyle \frac{\lambda^2 \pi^2}{4} \epsilon^{\alpha \nu \rho \sigma } \mathcal{B}_\nu \Tr{T_k^\mp L_\rho L_\sigma}
\end{equation}
where the upper (lower) sign corresponds to $V (A)$. In addition, $T_k^\mp \equiv iU^\dagger \left[\frac{\tau_k}{2},U \right]_\mp$, and $[\cdot,\cdot]_\mp$ refer to the commutator and the anti-commutator, respectively. The weak neutral current is then ($g\sin \theta_W = e$)
\begin{equation} \label{ec:JNCstructure}
    J_{NC}^\nu =- \frac{g}{c_W} \left[\frac{1}{2} \left(J_{3,V}^\nu - J_{3,A}^\nu \right) - s^2_W J_{EM}^\nu \right] .
\end{equation}
For the electromagnetic current, only the vector current contribution can be derived in this way. An additional baryon current density component originates from the Wess-Zumino-Witten term \cite{Callan:1983nx}, which is induced by QCD anomalies and must be incorporated into the model. However, for our purposes, it is sufficient to use the Gell-Mann–Nishijima formula, leading to
\begin{equation}
    J_{EM}^\nu = J^\nu_{3,V} + \frac{1}{2} \mathcal{B}^\nu .
\end{equation}

\prlsec{Semi-classical quantization}
The classical skyrmion solutions only provide the mass and atomic weight number (=baryon number) for the nucleus they are supposed to describe. For a description of the spin and the number of protons $Z$ and neutrons $N$, which are good quantum numbers of nuclei, the corresponding rotational and isorotational degrees of freedom must be introduced and quantized. Here, isospin is related to $Z$ and $N$ via $2i_3 = Z-N$, where $\hat{I}_3$ is the third component of the isospin operator, and $i_3$ its eigenvalue for a nuclear state.

In the Skyrme model this is achieved by a well-known procedure called rigid rotor quantization. In a first step, time-dependent rotations and isorotations are introduced into a static skyrmion $U_0(\mathbf{x}) $ via $U(t,\mathbf{x}) = A(t) U_0 (R(t) \mathbf{x}) A^\dagger(t)$,
where $ {A(t) \in SU(2)}$ and ${R(t) \in SO(3)}$.
When this expression is inserted into the Lagrangian, 
the terms without time derivatives just add up to minus the skyrmion mass, because the (iso)rotations are symmetries. The time-derivative terms, on the other hand, lead to a quadratic form in angular velocities. In a next step, this rigid-rotor Lagrangian is transformed to an equivalent Hamiltonian by a Legendre transformation, and the angular velocities are replaced by their conjugate momenta, the body-fixed spin ($\mathbf{L}$) and isospin ($\mathbf{K}$) angular momenta. In addition to introducing the rotational d.o.f., the matrix $R$ and its iso-rotational equivalent $(R_A)_{ij}= \tfrac{1}{2} \Tr (\tau_i A \tau_j A^\dagger) $ play a second role by providing the transformations from the body-fixed to the space-fixed spin ($\mathbf{J}$) and isospin ($\mathbf{I}$) angular momenta via 
$\mathbf{I} = - R_A\, \mathbf{K}$ and $\mathbf{J}=- R^T \mathbf{L}$, which implies $\mathbf{I}^2 = \mathbf{K}^2$ and $\mathbf{J}^2 = \mathbf{L}^2$. 

The angular momenta $\mathbf{I},\mathbf{J},\mathbf{K},\mathbf{L}$ are promoted to quantum operators $\hat{\mathbf{I}}$, etc.,  by imposing the usual angular momentum commutation relations. Nuclear states in this space are characterized by the corresponding eigenvalues,
$ \ket{\Psi} = \ket{ii_3 k_3} \otimes \ket{j j_3 l_3}$. The currents discussed above and their time components, which are the relevant ones for the form factors, become non-trivial operators if they contain time derivatives of the Skyrme field. This is not the case for the baryon current, which acts trivially (proportional to the identity) in this space. 
The components $J^0_{k,V(A)}$, on the other hand, are linear in time derivatives and therefore become linear operators in $\hat{\mathbf{I}}$, etc. 

For the BPS Skyrme model and for the axially symmetric ansatz \eqref{ec:nBPS2}, the vector current operator $J^0_{3,V}$ and its expectation value w.r.t. $ \ket{\Psi}$ have already been calculated in \cite{Adam:2013wya}, see eq. (18) of that paper (a much more detailed calculation can be found in \cite{naya}). The result is 
\begin{eqnarray} \label{ec:J0V}
\langle \hat{J}^0_{3,V} \rangle
  &=& 4\pi^4\frac{\lambda^2i_3}{\mathcal{I}_3}r^2 (\mathcal{B}^0(r) )^2 \frac{1+B^{-2}\cos^2\theta}{3B^2+1}\nonumber \\
   &=&  \frac{\lambda^2i_3}{\mathcal{I}_3 r^2} \xi_r^2 \sin^4 \xi \frac{B^2+\cos^2 \theta}{3B^2+1}.
\end{eqnarray} 
This is approximately spherically symmetric for sufficiently large $B$,  where $\frac{B^2+\cos^2 \theta}{3B^2+1} \sim \frac{1}{3}$.
Further, $\mathcal{I}_3$ is the 3-3 component of the isospin moment-of-inertia tensor, which is the integral of a known expression of $\xi$ (given in \eqref{eq:exactsol}) and its derivative $\xi_r$. For our choice of potential, the integral is analytic and given by:
\begin{equation}
    \mathcal{I}_3 = \frac{64 \, \pi}{105} \frac{\mu^2}{B^2}R_B^5.
\end{equation}
In principle, we are still missing the expectation value of the axial-vector current operator. However, it turns out that, for the axially symmetric ansatz, $ \langle \hat{J}_{3,A}^0 \rangle = 0$. This result agrees with the fact that the axial-vector contribution vanishes for $0^+ \rightarrow 0^+$ transitions \cite{Giomataris:2005fx}. 

Thus, for the charge densities of our interest, we get 
\begin{equation} \label{ec:J0EM}
    \langle \hat{J}^0_{\rm EM} \rangle  =  \frac{1}{2} \mathcal{B}_0 +  \langle \hat{J}^0_{3,V} \rangle,
\end{equation}
and 
\begin{equation} \label{ec:J0NC}
     \langle \hat{J}_{\rm NC}^0 \rangle  =   \frac{g}{c_W} \left( \frac{1}{2} s_W^2 \mathcal{B}_0 - \left(\frac{1}{2}-s_W^2 \right) \langle \hat{J}_{3,V}^0 \rangle \right) .
\end{equation}

\prlsec{Computing the form factors}
FF take into account the extended character of nuclei in scattering and other interaction processes. In the Breit frame, where only momentum $\mathbf{q}$ but no energy is transferred from the nucleus to its scattering partner (electron or neutrino), the FF is the purely spatial Fourier transform of the relevant charge distribution ${\mathcal{Q}}_\rho \sim \langle \hat{J}^0_\rho \rangle$ \cite{Karliner:2015qoa}:
\begin{equation} \hspace*{-0.4cm}
    F_\rho (\mathbf{q})=\int \frac{\mathcal{Q}_\rho(\mathbf{x})}{Q_\rho} e^{-i \mathbf{q} \cdot \mathbf{x}} d^3 \mathbf{x} \approx \int \frac{4\pi \mathcal{Q}_\rho(r)}{Q_\rho} \frac{\sin qr}{qr} r^2dr 
\end{equation}
where $\rho$ above denotes the electromagnetic current (EM) or weak neutral current (NC), respectively.
In the last equation, we assume an approximate spherical symmetry of $\mathcal{Q}_\rho$. Furthermore, the FF is normalized to $F(q=0)=1$ by convention, with the normalization factor being the associated total charge,
$
    Q_\rho=4\pi \int_0^\infty r^2\mathcal{Q}_\rho(r)dr
$.

Since only the small momentum transfer region is experimentally accessible for weak charge FF, it is convenient to expand the weak FF about $q=0$ in terms of the even moments of the density distribution,
\begin{equation} \label{ec:taylor}
    F_\rho(q) =  \sum_{n=0}^\infty \frac{(-1)^n}{(1+2n)!} \langle R^{2n}_{\rho} \rangle(B,i_3) q^{2n},
\end{equation}
with
\begin{equation}
        \expval{R^{2n}_\rho} = \frac{4\pi}{Q_\rho}\int r^{2n+2}\mathcal{Q}_\rho(r)dr.
\end{equation}
In the BPS Skyrme model, the vector density $\langle \hat{J}^0_{3,V} \rangle $ is proportional to $r^2(\mathcal{B}^0(r))^2$, (see \eqref{ec:J0V}), hence we can write the moments of the EW charge densities in terms of the moments of $\mathcal{B}^0$ and $(\mathcal{B}^0)^2$:
\begin{eqnarray}
    \int r^{2n+2}  (r\mathcal{B}^0)^2 ~dr = \frac{4\mu^2}{\pi^4 \lambda^2} \frac{R_B^{2n+5}}{(2n+5)(2n+7)},\\[2mm] \int r^{2n+2}  \mathcal{B}^0~dr = \frac{\mu}{2\sqrt{2} \lambda \pi^{3/2}} \frac{\Gamma(n+3/2)}{\Gamma(n+3)} R_B^{2n+3},
\end{eqnarray}
where $\Gamma$ represents Euler's gamma function. For instance, the weak neutral current FF $F_{NC} (q)$  for \emph{any} given nucleus with quantum numbers $(B,i_3)$ leads to the moments
\begin{equation}
     \displaystyle \langle R^{2n}_{\rm NC} \rangle  \!=  \! \displaystyle \frac{4 \pi  R_B^{2n}}{Q_{\rm NC}}\left[
     \frac{s^2_W}{2}\frac{B\Gamma(n+3/2)}{\pi^{3/2}\Gamma(n+3)}   \!+\!  \displaystyle  
      \frac{35 \qty( s^2_W-1/2)i_3}{4\pi (2n+5)(2n+7)}  \right],
\end{equation}
where $Q_{\rm NC} = \frac{1}{2} (s^2_W B - (1-2s^2_W)i_3)$. Similarly, for the electromagnetic charge form factor, we get
\begin{equation}
    \langle R^{2n}_{\rm EM} \rangle=   \displaystyle \frac{4 \pi  R_B^{2n}}{Q_{\rm EM}}\left[
     \frac{B\Gamma(n+3/2)}{2\pi^{3/2}\Gamma(n+3)} +
      \frac{35 i_3}{4\pi (2n+5)(2n+7)}  \right],
\end{equation}
with $Q_{\rm EM} = \frac{B}{2} + i_3$. In both types of interaction, the analytical approximation in \eqref{ec:taylor}, upon inclusion of the corresponding EM or NC moments, shows very good agreement with the exact solution in the region considered, assuming terms up to $n = 10$.

Finally, $R_{B}$ depends solely on the values
of the free parameters $\lambda$ and $\mu$, which are fitted to nuclear masses and  radii (see \cref{Appendix A}). With our choice of values, we have 
\begin{equation} \label{R-B}
    R_{B}= 1.3626 \, B^{1/3} \,\fm .
\end{equation}

\begin{figure}
    \centering
\includegraphics[width=1\linewidth]{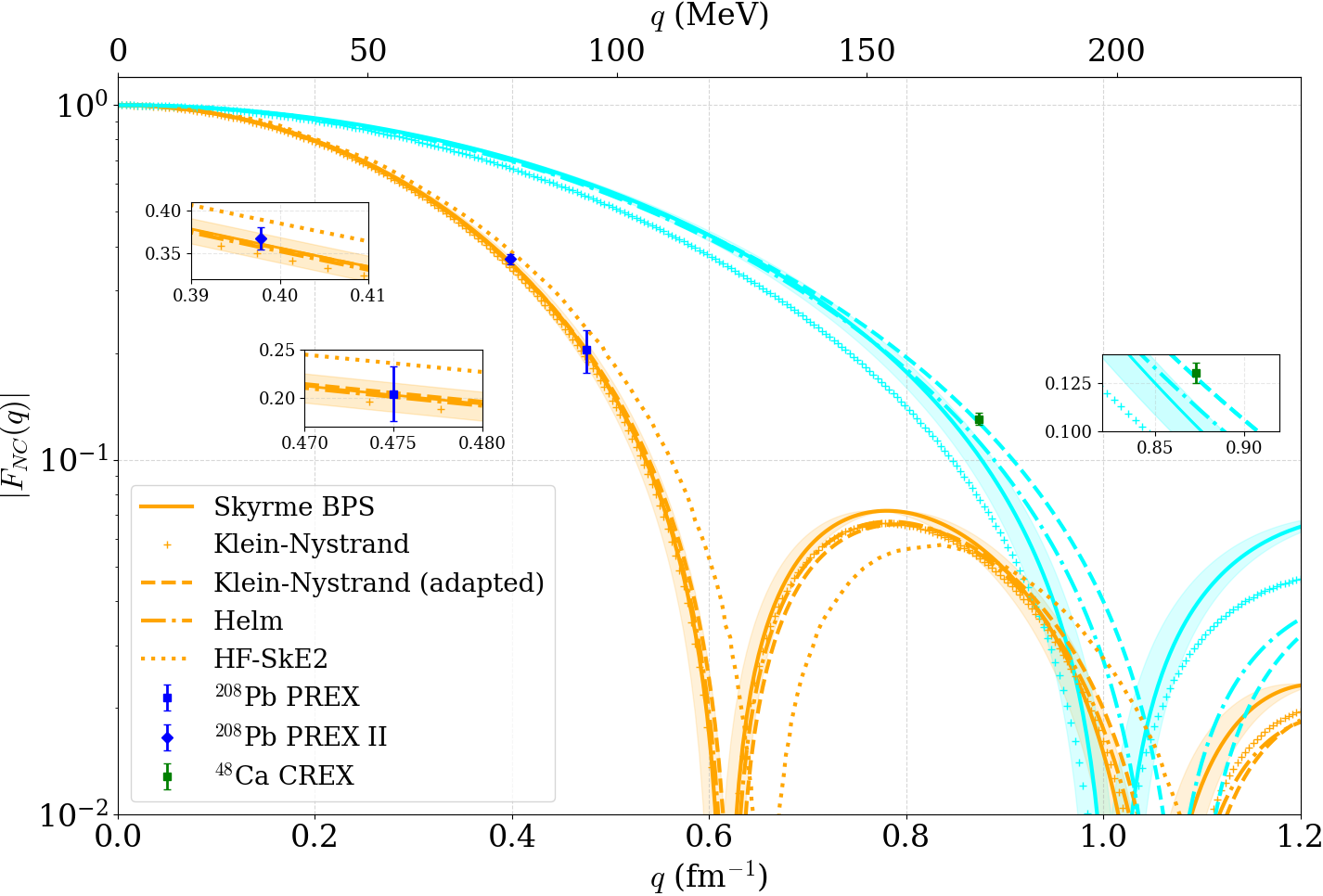}
    \caption{ Neutral current data for the $^{208}\rm Pb$ and $^{48}\rm Ca$ form factors
    \cite{Horowitz:2012tj,PREX:2021umo,CREX:2022kgg}.
    Solid lines represent the BPS Skyrme model
    prediction, and the band accounts for the uncertainty in the main model parameter $R_B$, which arises from different choices of the radii-mass region to fit (\cref{Appendix A}). 
    For comparison, we show the Helm (dot-dashed), KN (crosses) and KNa (dashed) phenomenological models, as well as the selected microscopic HF-SkE2 model (dotted), for illustration.}
    \label{form-nc}
\end{figure}

\begin{figure}
    \centering
\includegraphics[width=1\linewidth]{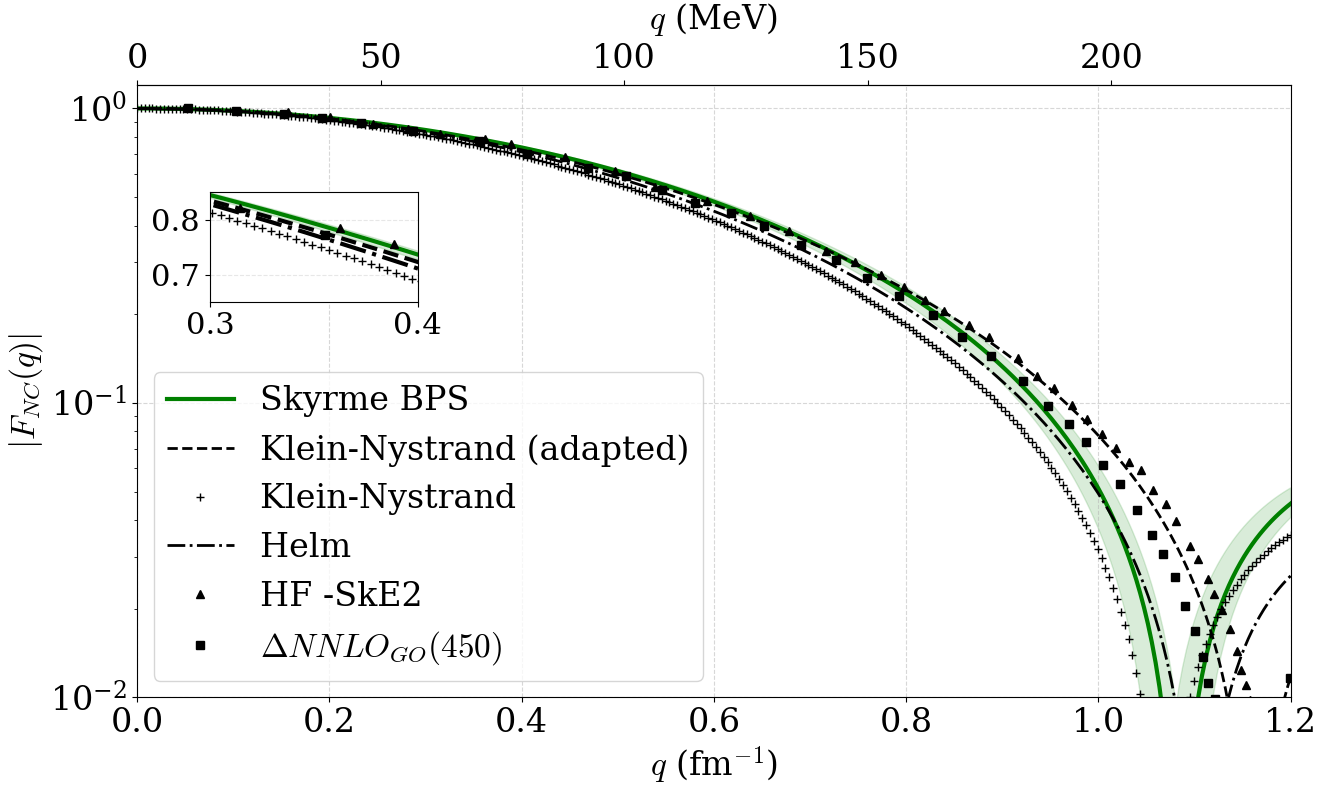}
    \caption{ Predicted neutral current form factors for $^{40}\rm Ar$. Solid lines represent the BPS model prediction, phenomenological models are included such as KN (crosses), KNa (dashed) and Helm (dot-dashed), and the two illustrative microscopic models HF-SkE2 (triangles) and $\Delta NNLO_{GO}(450) $ (boxes). }  
    \label{form-ar}
\end{figure}

As shown in appendix A, an excellent agreement is found when comparing the above formulas with measured EM FF up to $q=140-175$MeV, the highest momentum corresponding to the lightest nucleus in this survey (${}^{40}$Ar). With this in mind, it is possible to explore the implications of the model for the weak neutral FF: 
\cref{form-nc} shows its value for $^{208}\rm Pb$ and $^{48}\rm Ca$, where experimental results are available from recent PVES experiments. We also compare with some
widely used phenomenological descriptions (Klein-Nystrand (KN): crosses, KN-adapted (KNa): dashed, Helm: dash-dotted) as well as a selected microscopic model (HF-SkE2), which have been employed in recent studies. The model-dependent bands are derived from different fitting assumptions in eq. 27 (see appendix A). It is seen that the agreement is very good for the measured FF-values in ${}^{208}$Pb, showing by contrast a 3-$\sigma$ deviation for ${}^{48}$Ca, which has been nevertheless measured around the maximum momentum for which the BPS-calculation can be expected to lose validity, based on the EM-FF analysis presented in appendix A. Details on their implementation can be found in \cref{Appendix B}. 
To further illustrate the potential of the BPS-formalism,
\cref{form-ar} presents the predicted weak neutral FF for $^{40}\rm Ar$. This nucleus is very important for CE$\nu$NS experiments, therefore several theoretical studies and phenomenological fits are already available \cite{COHERENT:2019iyj}. Currently, there exists no experimental PVES data, so the results of \cref{form-ar} can be viewed as genuine predictions of the different models. For details see \cref{Appendix B}.

\prlsec{Conclusions}
In this work, we have explored the electroweak
currents and the corresponding form factors for
large-mass nuclei (above or around $B=40$), which are experimentally relevant
to the study of coherent elastic neutrino-nucleus
scattering (CE$\nu$NS) and parity-violating electron scattering
(PVES) experiments, from the point of view of
the generalized Skyrme model, in which nuclei are modelled as
topological solitons. In this approach, the computation
of spatial densities for electroweak charges is reduced to
finding the classical soliton configuration. 

We restricted our calculation to the Bogomolnyi-Prasad-Sommerfield 
term under the pion mass potential,  which constitutes the leading chiral order contribution to the potential. This sub-model  has shown its prowess in the description of high-density baryonic systems,
and it enables the classical soliton solution to be obtained analytically. 
The model contains two parameters which combine effectively into a single global 
one, $R_B$ (root-mean-square charge radius), on which FF analytically depend. This represents, arguably, the simplest solution of the generalized Skyrme model and, yet,
the calculated electromagnetic form factors display an excellent agreement with the experimental data for momentum transfers of up to  $\sim$ 140-175\,MeV. 

For $^{208}\rm Pb$ and $^{48}\rm Ca$, where 
experimental data for the weak neutral FF
exist, our model better reproduces observations than the widely used Klein-Nystrand (KN) model, and it is generally comparable with other models (and within the dispersion between them) up to the aforementioned momentum range. Yet, predictions depend on a single parameter, effectively $R_B \sim (\lambda/\mu)^{1/3} = 1.3626$ fm. 

Our work can be interpreted on the one hand as an effective field theory
calculation of the electromagnetic and weak FF of nuclei. On the other hand, it offers a simple single-parameter analytical formula for them, with the free parameter completely fixed by the electromagnetic sector. As data keeps accumulating on CE$\nu$NS and PVES scattering experiments, a first-principle global description with a minimum of free parameters is called to be a powerful asset to identify trends or anomalies, that could point to Beyond Standard Model physics. Furthermore, whereas momentum transfers in current PVES data are in some cases at the limit of applicability of the BPS-model, past and near-future CE$\nu$NS experiments deal with much more convenient values of few 10's of MeV.
\\

\prlsec{Acknowledgements}
The authors acknowledge financial support from the Spanish Research State Agency under project PID2023-152762NB-I00, the Xunta de Galicia under the project ED431F 2023/10 and the CIGUS Network of Research Centres, the María de Maeztu grant CEX2023-001318-M funded by MICIU/AEI /10.13039/501100011033, and the European Union ERDF. The work of A.X.L.F. was also supported by grant ED481A-2025 (Consellería de Cultura, Educación, Formación Profesional y Universidades, Xunta de Galicia). 
 A.G.M.C. is supported by grants No. ED481B-2025/059 and ED431B-2024/42 (Consellería de Cultura, Educación, Formación Profesional y Universidades, Xunta de Galicia).
\bibliography{biblio}


\appendix
\section{Fitting the free parameters}
\label{Appendix A}
To fix the parameter values of $\lambda$ and $\mu$ from the BPS lagrangian density, we fitted the classical skyrmion energy to the empirical nuclear masses as a function of the baryon number~$B$,
\begin{equation}
    E_{BPS}=\frac{64 \sqrt{2} \pi}{15} \mu\lambda B,
\end{equation}
obtaining the parameter $ \mu \lambda$, and the root-mean-square charge radii of the BPS skyrmions as a function of the nuclear isospin and the baryon number, which allowed us to extract the parameter $\left(\frac{\mu}{\lambda}\right)^{1 / 3} $
\begin{equation}
  \sqrt{ \left\langle r^2\right\rangle} =\sqrt{\frac{\int r^2 \left \langle J^0_{EM}  \right\rangle d V}{\int  \left \langle J^0_{EM}  \right \rangle d V}} = \left(\frac{\lambda}{\mu}\right)^\frac{1}{3} \sqrt{\frac{B^{2/3} \left(20\, i_3 + 9 B\right)}{\left(18 \, i_3 + 9 B\right) }}.
\end{equation}
The experimental dataset consists of all stable nuclei with even mass number from ${}^{40}$Ar to ${}^{208}$Pb, except for ${}^{156}$Dy, ${}^{158}$Dy and ${}^{160}$Dy that were removed due to their large experimental uncertainties.
We have used a least mean squares fit with the Trust Region Reflective (TRF) algorithm \cite{Scipy}, but we 
weighted the contributions of individual nuclei by their uncertainties only for the nuclear charge radius fit, since this is the  experimental input relevant for the form factors. 
The values for the parameters after fitting are 
\begin{align} \label{ec:lambdamu}
    \lambda &=  6.62732 \pm 0.00045 ~\text{MeV}^{1/2}\text{fm}^{3/2}, \notag\\  \mu &= 7.40914 \pm 0.00050 ~\text{MeV}^{1/2}\text{fm}^{-3/2},
\end{align}
leading to \cref{R-B} and to
  $E_{BPS}= 931.0\, B\, \MeV $. 
However, the parameter uncertainties obtained from Eq. \eqref{ec:lambdamu} do not incorporate possible systematic uncertainties of the model. 
In order to obtain an estimate of the variation of the form factor arising from the choice of fitted nuclei, we have shaded the band spanned by two extreme selections of nuclei used in the fit: one fit performed using nuclei with baryon number between 40 and 84, and a second fit including only nuclei with baryon number larger than 182.

\begin{figure}[h]
    \centering
\includegraphics[width=0.9\linewidth]{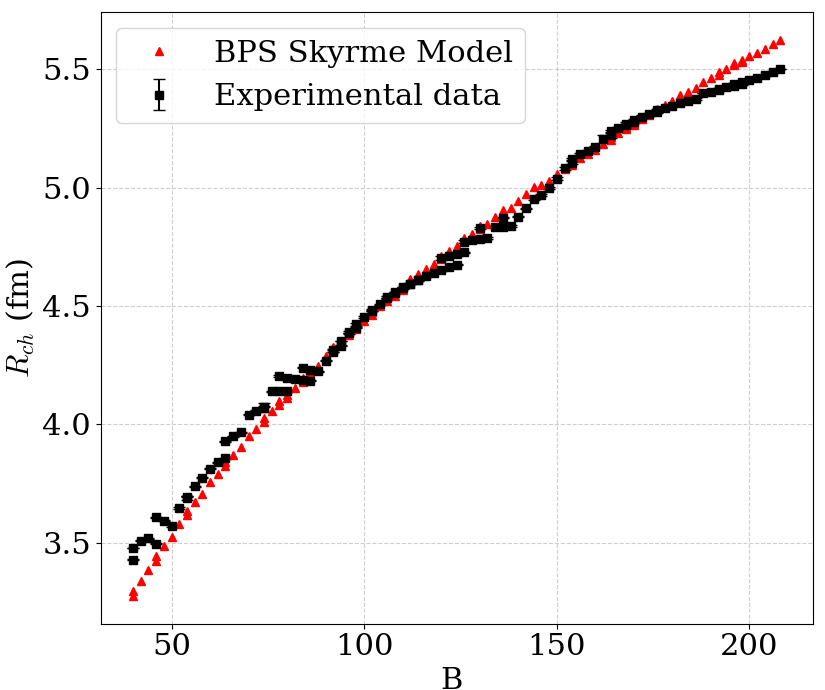}
    \caption{Experimental data of electromagnetic charge radii (black squares) for the nuclei used to fit the effective parameter $R_B$, and their predicted values using the BPS Skyrme model with the fitting of \eqref{ec:lambdamu} (red triangles). Experimental data are from \cite{IAEA}.}
    \label{fig:fit}
\end{figure}

\begin{figure}[h]
        \centering
\includegraphics[width=1\linewidth]{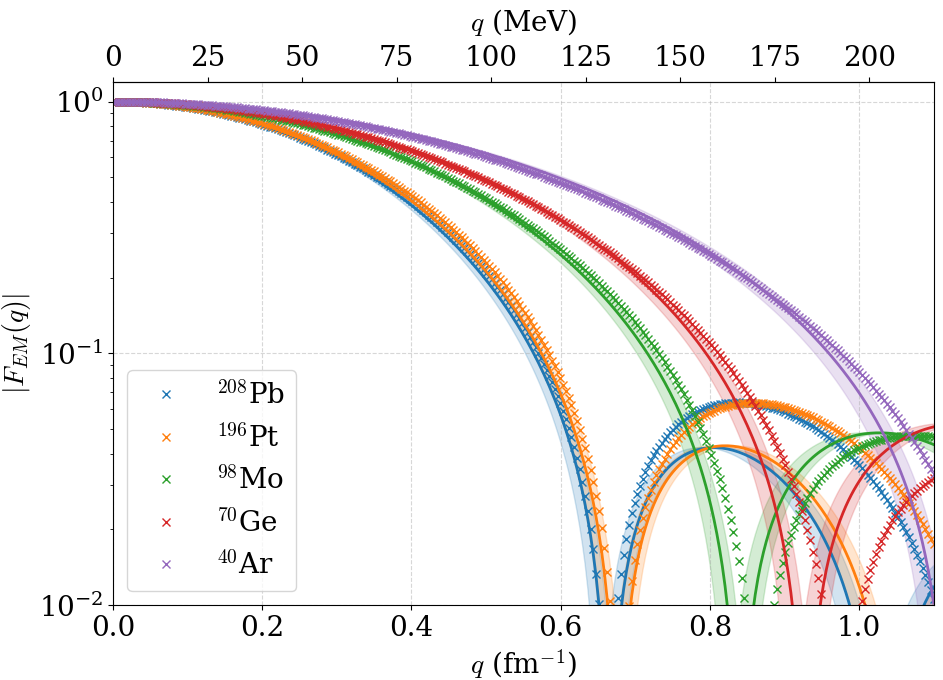}
        \caption{Electromagnetic form factors for different nuclei. Solid lines represent the prediction of the BPS model,
         with the band accounting for different choices of the main model parameter $R_B$, compatible with measured nuclear radii in different mass ranges. Crosses correspond to experimental values obtained
        from a Fourier–Bessel fit to the elastic electron scattering of \cite{DeVries:1987atn}}.
        \label{form-em}                        
    \end{figure}

For the central value (fit to all nuclei), we show the result of the fit in \cref{fig:fit}. We see that the fit is quite accurate, especially in the region of nuclear mass number $ 80 \le B \le 180$. In particular, the relative error of the RMS radius is $0.135\%$. We also show the resulting electromagnetic form factors for a representative set of medium ($B\simeq 40$) to large ($B\simeq 200$) nuclei in \cref{form-em}. They agree with the experimental values with excellent precision up to values of $q$ which, depending on the nuclei, vary between $q\simeq 140-175\, \MeV$. 

\section{Comparison with other models}
\label{Appendix B}
Here we briefly describe the different FF that we compare with our results in \cref{form-nc} and in \cref{form-ar}.
The Klein-Nystrand FF \cite{Klein:1999qj}  is obtained from the convolution of a short-range Yukawa potential with a hard-sphere distribution:
\begin{equation}
    F_{\mathrm{KN}}(q)=
    3\,\frac{
    \sin \left(q R_A\right)-q R_A \cos \left(q R_A\right)
    }{\left(q R_A\right)^3}
    \left[1+\left(q a_k\right)^2\right]^{-1}.
\end{equation}
 This is the FF adopted by the COHERENT collaboration \cite{COHERENT:2017ipa}. The parameter $a_k$ determining the range of the Yukawa potential is commonly fixed to $a_k = 0.7~\text{fm}$ \cite{Klein:1999qj, Davies1976}, but the nuclear radius $R_A$ admits two different choices. The first, global to all nuclei, consists of expressing $R_A = r_0 A^{1/3}$ and fitting $r_0 = 1.23~\text{fm}$ as a global parameter.  In the second, known as the \textit{adapted} choice, $R_A$ is chosen so that the root-mean-square radius of the Klein-Nystrand distribution matches the  weak radius of the nucleus under study \cite{VanDessel:2020epd}. As this weak radius is usually extracted from the weak form factor, this is essentially a direct fit to that form factor, at least in cases where only one data point is available (e.g., ${}^{48}$Ca).

To obtain the Helm Form Factor \cite{ENGEL1992} it is assumed that the nucleonic distribution is given by  a convolution of an uniform density with radius $R_0$ and a Gaussian profile characterized by a parameter $s$:
\begin{equation}
	F_{\mathrm{H}}\left(q\right)=3 \frac{\sin \left(q R_0\right)-q R_0 \cos \left(q R_0\right)}{(q 	R_0)^3} e^{-q^2 s^2 / 2},
\end{equation}
The parameter $s$ is fixed to $s=0.9$ fm and $R_0$ is again fixed in an adapted manner, so that the root-mean-square radius of the form factor matches the weak radius of the nucleus \cite{Sierra2019}. 

The adapted Klein-Nystrand and Helm models share the drawback of requiring a weak radius, typically extracted from weak form factor measurements and therefore model dependent, which limits their predictive power. The weak radius of ${}^{208}$Pb is taken from \cite{Horowitz:2012tj}. The value for ${}^{48}$Ca is extracted from \cite{CREX:2022kgg}, assuming $R^{\,{}^{48}\mathrm{Ca}}_{\mathrm{em}} = 3.4771(20)$ fm. For ${}^{40}$Ar, the radius has been taken from the Skyrme–Hartree–Fock predictions reported in \cite{AbdelKhaleq:2024hir}.

Moreover, we included two microscopic models, illustratively. For the $^{208}\rm Pb$ and $^{40}\rm Ar$ we show the FF obtained from the densities predicted by the HF-SkE2 model \cite{VanDessel:2020epd},  a microscopic many-body nuclear theory model where the nuclear ground
state is described in a Hartree–Fock (HF) approach with a SkE2 nuclear potential. This model has shown excellent agreement with the relativistic mean field theory (RMF)
predictions \cite{Yang:2019pbx} for $q \lesssim 1.8 \mathrm{fm}^{-1} $ \cite{VanDessel:2020epd}. In fact, the curves representing HF-SkE2 for ${}^{208}$Pb and ${}^{40}$Ar in \cref{form-nc} and \cref{form-ar}, 
can be interpreted as coming from RMF calculation, within the size of the markers.
For the case of $^{40}\rm Ar$, we also included \cite{Payne:2019wvy}, where a nuclear hamiltonian derived from chiral effective field theory is solved using nuclear coupled cluster theory.

\end{document}